\documentclass[aps,prb,citeautoscript,reprint,superscriptaddress,groupedaddress,amsmath,amssymb]{revtex4-2}

\usepackage{graphicx}
\usepackage{dcolumn}
\usepackage{bm}
\usepackage{hyperref}

\newcommand*{\citen}[1]{%
  \begingroup
    \romannumeral-`\x 
    \setcitestyle{numbers}%
    \cite{#1}%
  \endgroup   
}

\usepackage{xcolor}

\begin{document}
\setcitestyle{super}

\title{V$_2$C-based lithium batteries: The influence of magnetic phase and Hubbard interaction}

\author{Jhon W. Gonz\'alez}
\email{jhon.gonzalez@usm.cl}
\affiliation{Departamento de F\'{i}sica, Universidad 
T\'{e}cnica Federico Santa Mar\'{i}a, Casilla Postal 
110V, Valpara\'{i}so, Chile.}

\author{Sanber Vizcaya}
\affiliation{Departamento de F\'{i}sica, Universidad 
T\'{e}cnica Federico Santa Mar\'{i}a, Casilla Postal 
110V, Valpara\'{i}so, Chile.}

\author{Eric Su\'arez Morell}
\email{eric.suarez@usm.cl}
\affiliation{Departamento de F\'{i}sica, Universidad 
T\'{e}cnica Federico Santa Mar\'{i}a, Casilla Postal 
110V, Valpara\'{i}so, Chile.}

\date{\today}

\begin{abstract}
MXenes are a family of two-dimensional materials that could be attractive for use as electrodes in lithium batteries due to their high specific capacity. For this purpose, it is necessary to evaluate magnitudes such as the lithium adsorption energy and the magnitude of the open-circuit voltage for different lithium concentrations. In this paper, we show through first principles calculations that in a V$_2$C monolayer we must consider the high correlation between the electrons belonging to vanadium to obtain correct results of these quantities. We include this correlation employing the Hubbard coupling parameter obtained by a linear response method. We found that the system is antiferromagnetic and that the quantities studied depend on the magnetic phase considered. Indirectly, experimental results could validate the theoretical value of the theoretical Hubbard parameter.

\end{abstract}

\maketitle

\section{\label{sec:intro}Introduction}
The family of 2D transition-metal carbides, carbonitrides, and nitrides collectively referred to as MXenes is another emerging family of 2D materials that has received growing interest due to their interesting physical and chemical properties.
Among the potential applications of this family of materials are their use as 
the basis for the manufacture of electrodes for lithium 
batteries \cite{Mishra_2021,SUN201480}.
The feasibility of materials as ion battery electrodes is characterized 
by the charge capacity --related to the energy stored per unit formula-- and 
by the open-circuit voltage --related to the change in the chemical potential 
of the electrode as a function of ion concentration--.
In particular, the V$_2$C is a promising 2D material for Li-ion electrodes 
due to its high theoretical specific capacity, estimated in $940$ mAh/gr. 
The V$_2$C capacity is higher than other members of the MXenes family and carbon-based electrodes\cite{Sato556}. 

However, the magnetic character of V$_{2}$C ground state is a matter of debate.
Due to the strongly localized nature of the d-electrons in transition metal atoms, 
the compounds involving such elements like vanadium are challenging to model 
using ab-initio techniques. Electron correlation effects may lead to parameter-dependent atomic structures, electronic band-gap and magnetic phases\cite{BAE2021100118}.
In some calculations the ground state phase is considered non-magnetic\cite{V2N_NM} while in others the authors claim it is antiferromagnetic\cite{V2N_AFM}.
The matter is not trivial, and as we will show the calculations with different magnetic phases lead to different results in the quantities 
associated with battery performance\cite{bennett2014influence}. 
The electronic correlation effects on the vanadium atoms are strong, an when they are considered in ab-initio calculations physical properties such as lattice constant, inter-atomic distances, and adsorption properties are modified. It impacts also on the magnetic phase of the ground state of the system. 

In this manuscript we evaluate, using first-principles calculations, the impact of the U-Hubbard (DFT+U) term in the electro-chemical properties of the monolayer V$_2$C as an electrode in a Lithium-ion based battery. 
We calculated first the U-Hubbard value for the vanadium atoms in V$_2$C by using the linear response method\cite{cococcioni2005linear}. The obtained value depends on the employed density functional theory (DFT) code. We obtained a value of 5.0 eV for {\sc Quantum-ESPRESSO}  and 4.1 eV for VASP. 
However, despite the difference in the values, we found the same stability order, similar magnetic moments, and charges, and  in particular similar ferromagnetic-antiferromagnetic energy differences in both codes. 

\begin{figure}[!hb]
\includegraphics[clip,width=0.49\textwidth,angle=0,clip]{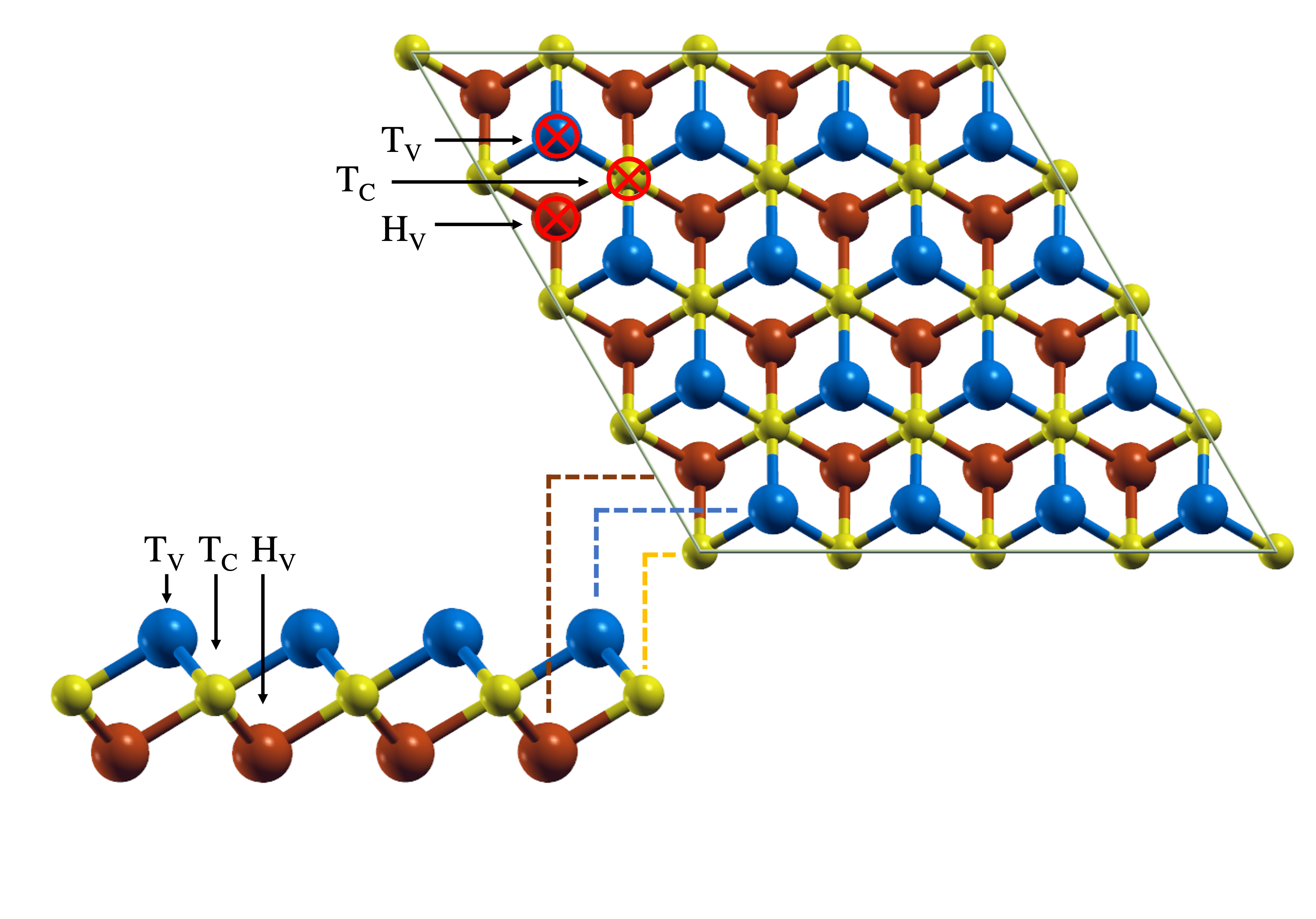} 
\caption{Schematic view showing top and side view. The lithium adsorption sites are
labeled.}
\label{Fig:scheme}
\end{figure}


Finally, using the U value obtained, we calculate the adsorption energy and the open-circuit voltage (OCV) in a 4x4 supercell as a function of the lithium concentration. We study the different adsorption sites of lithium atoms in V$_2$C and find that regardless of the magnetic character of the V$_2$C, the lithium tends to be adsorbed on top of the carbon atom.
Our results show agreement with recent experimental results and validate the use of the U parameter to obtain the electrochemical properties of mxenes.\cite{wu2020synthesis}.
In that sense, from measurable, battery-related properties, the magnetic state of the V$_2$C could be inferred and used as an indirect validation of the U-Hubbard parameter determined by theoretical methods.

\section{Methodology}
\label{metod}
The systems were modeled within the density functional theory (DFT) using the plane-wave approximation implemented in VASP and {\sc Quantum-ESPRESSO} (QE). In both codes, we used the generalized  gradient approximation (GGA) proposed by Perdew-Burke-Ernzerhof (PBE) \cite{PBE}, and we  included the van der Waals correction using the DFT-D3 correction\cite{grimme2010consistent}. 

We performed a thorough calculation, with and without the U-Hubbard term. 
We compute the U-Hubbard term using the linear response approach\cite{cococcioni2005linear} and compare it with previous reports\cite{Ya_Meng_Li_2020,Nyandelger_2020,ji2015different,hu2014investigations}.
The U-Hubbard term is the computationally cheapest addition that can be used to  include correlation effects of electrons 
in d and f orbitals, and its use allows a better fit of theory to experimental measurements. 
However, the specific U-Hubbard value must often be determined empirically only when 
adequate experimental data (electronic bandgap, magnetic character, and others) are available\cite{xu2015linear}. 
The linear response method determines the optimal value using the differences 
between the derivative of the screened and bare energies with respect to 
the occupation $N_i$ of the n-localized states (d-orbitals) at the atomic site
$i$ upon a perturbation $\alpha$, as
follows\cite{cococcioni2005linear,kulik2006density,floris2011vibrational}:
 \begin{equation} \label{eq:U} 
 U=\left( \frac{ \partial N_i^{SCF}}{\partial \alpha_i} \right)^{-1} -\left( \frac{ \partial N_i^{NSCF}}{\partial \alpha_i} \right)^{-1}.
\end{equation} 
In practical terms, the linear-response methodology involves a small perturbation $\alpha_i$ to the atomic site $i$, acting as a Lagrange multiplier to maximize the functional to the d-orbitals belonging to one of the vanadium atoms to excite charge fluctuations in these orbitals and solve the self-consistent Kohn-Sham equations to obtain perturbed occupations \cite{Xu_U}.

For vanadium atoms in V$_2$C-monolayer, in {\sc Quantum-ESPRESSO} we use  the Lowdin-orthogonalized atomic wave-functions (ortho-atomic option), we find a linear-response  
$U_{\mathrm{QE}}= 5.0$ eV. Whereas, with VASP, we find a  linear-response value of $U_{\mathrm{VASP}} = 4.1$ eV. Despite the differences between the two implementations, there is an agreement between the calculated quantities: relative energies, stability order, structural properties, adsorption properties, charges, and magnetic moments.
This difference between $U_{\mathrm{QE}}$ and $U_{\mathrm{VASP}}$ is related to differences in the implementation of correlation effects in both codes. VASP uses the projectors of the pseudopotentials (PAW type mostly) to define the atomic occupations to be used in the Hubbard correction.  {\sc Quantum-ESPRESSO} atomic wave-functions are instead mostly used to project Kohn-Sham states.  An in-depth discussion of the differences in DFT+U implementation between VASP and QE can be found in ref. ~\citen{wang2016local}.

For the calculations with  {\sc Quantum-ESPRESSO}, we use a kinetic energy cutoff for the wave functions of 80 Ry 
($\sim 1090$ eV) and 650 Ry ($\sim 8844 $ eV) for charge density. In VASP calculations, 
the cutoff energy was set to 520 eV. In both codes, in the calculations to obtain the magnetic order, we use a dense $21 \times 21\times1$ k-grid. 
To perform Li-adsorption calculations, we extended the system to a $ {4}\times {4}$ 
supercell with a $3\times3\times1$ k-grid. 

In order to test a number of concentrations of lithium atoms in the V$_2$C monolayer, we need to work with a supercell. Based on previous works\cite{V2N_NM,V2N_AFM}, we extend our cell four times in the in-plane direction; we will refer to this cell as $4 \times 4$ cell.
The order of stability referred to the magnetic phase is the same in the primitive cell ($1 \times 1$) as in the  $4 \times 4$ supercell.
After relaxing the $4 \times 4$ cell we found a crystallographic phase with lower energy than the $1 \times 1$ primitive cell, This effect is especially noticeable when ferromagnetic solutions are considered. 

We calculated the average absorption energy ($E_{ads}$) of N lithium atoms in the 
$ {4}\times {4}$ supercell using:
 \begin{equation} \label{eq:Eads} 
E_{ads}= \frac{E_{Li_{x_1}V_2C} - E_{V_2C} -  N\, E_{Li} } {N},
\end{equation}
where $E_{V_2C}$ refers to the total energy of pristine V$_{2}$C layer, 
$E_{Li_{x_1}V_2C}$ is the energy of the system with N lithium atoms, that corresponds to
a concentration $x_i$ of lithium ions adsorbed on the V$_{2}$C monolayer, and $ E_{Li}$ refers to the 
chemical potential of a lithium ion. 
The value of $ E_{Li}$ is obtained by considering one Li atom 
as a free-ion (gas phase). With this definition, more negative values represent higher adsorption.

Another parameter used to assess the theoretical performance of lithium-ion batteries is the open-circuit voltage (OCV) profile. In practical terms, the open-circuit voltage is associated with the change of the 
chemical potential of the electrode as the lithium number changes, and it is related 
to the slope of the formation energy ($E_{form}=N E_{ads}$) as a function of lithium concentration. 
The OCV profile of an ion adsorption on a surface can be calculated from the change in the Gibbs free energy of the system. When neglecting the changes in volume and entropy, 
the OCV expression simplifies to a difference of total
energies\cite{akgencc2019two,Ya_Meng_Li_2020,zhou2004first}, as
\begin{equation} \label{eq:OCV} 
OCV \approx \frac{E_{Li_{x_2}V_2C} - E_{Li_{x_1}V_2C} - \left(x_2 - x_1 \right)E_{Li}  } {\left(x_2 - x_1 \right)e},
\end{equation}
where e is the electron charge, $E_{Li_{x_i}V_2C}$ is the total energy of the V$_2$C with
a concentration of $x_i$ lithium ions, and $E_{Li}$ is the chemical potential of 
lithium ion\cite{liu2021modulating}.
Experimentally $E_{Li}$ is a reference point used to measure the OCV in Lithium battery.  Lithium foils are usually used for that purpose as a reference electrode and also as auxiliary electrodes in experimental setups \cite{wu2020synthesis}. For that reason in our calculations we employed the $E_{Li}$ in a BCC crystal.

Another relevant parameter to characterize any active material as an electrode for batteries is the specific capacity, also called theoretical gravimetric reversible capacity\cite{ashton2016computational}. The specific capacity is the amount of charge stored in a material per mass unit. It is calculated from the chemical composition of the material storing the charge\cite{liu2021modulating}. 
The specific capacity indicates the amount of charge stored per unit weight of the formula. The experimental value is obtained from a  voltage-time curve in a galvanostatic cycle, and the theoretical specific capacity reads
\begin{equation} \label{eq:capacity} 
C = \frac{  N F } {M_w},
\end{equation} 
where, N is the number of ions, F is the Faraday constant, 
and $M_w$ is the molecular weight of the electrode.

\section{Results}
\subsection{Stability order}
First, we proceed to obtain the crystallographic phase of the V$_2$C monolayer; due to the magnetic nature of vanadium atoms, we must consider three magnetic phases, whether the system is non-magnetic (NM), ferromagnetic (FM), or antiferromagnetic (AF).
The V$_2$C monolayer has a crystallographic structure with a lattice similar to that of MoS$_2$ with an arrangement of 1T of the three atoms in the unit 
cell\cite{kosmider2013large,cortes2018stacking}.
However, in MXenes\cite{BAE2021100118} and some dichalcogenides\cite{CDW0,CDW1,CDW2}, the high electronic correlation distorts the unit cell, creating superstructures. This crystallographic phase is known as the charge density wave (CDW) phase due to the appearance in several of these structures of this type of collective excitation.
Such structural distortion is stronger in ferromagnetic solutions and yields finite-size effects.
First, we relax the 1T structure, and later from that cell, we build a $4 \times 4$ supercell. 
The relaxation of the FM solution resulted in a crystallographic phase 16.5 meV per formula 
unit lower in energy than the more symmetric 1T phase. We use this $4 \times 4$ superstructure 
to determine the magnetic phase and, subsequently, the study of the magnitudes associated 
with the performance of monolayer V$_2$C as a battery component.

Let us now study how the magnetic phase depends on the electronic correlation in the material; this effect is included through the parameter U of the Hubbard interaction.
The magnetic moment depends on the value of U. For $U=0$, the magnetic moment is zero, so
the ground state is non-magnetic, the solutions with magnetic moment (FM or AF) are 1 eV 
per formula unit above. It was necessary to force the magnetic solutions and for that reason 
we have not included them in the manuscript.

\begin{figure}[!ht]
\includegraphics[clip,width=0.45\textwidth,angle=0,clip]{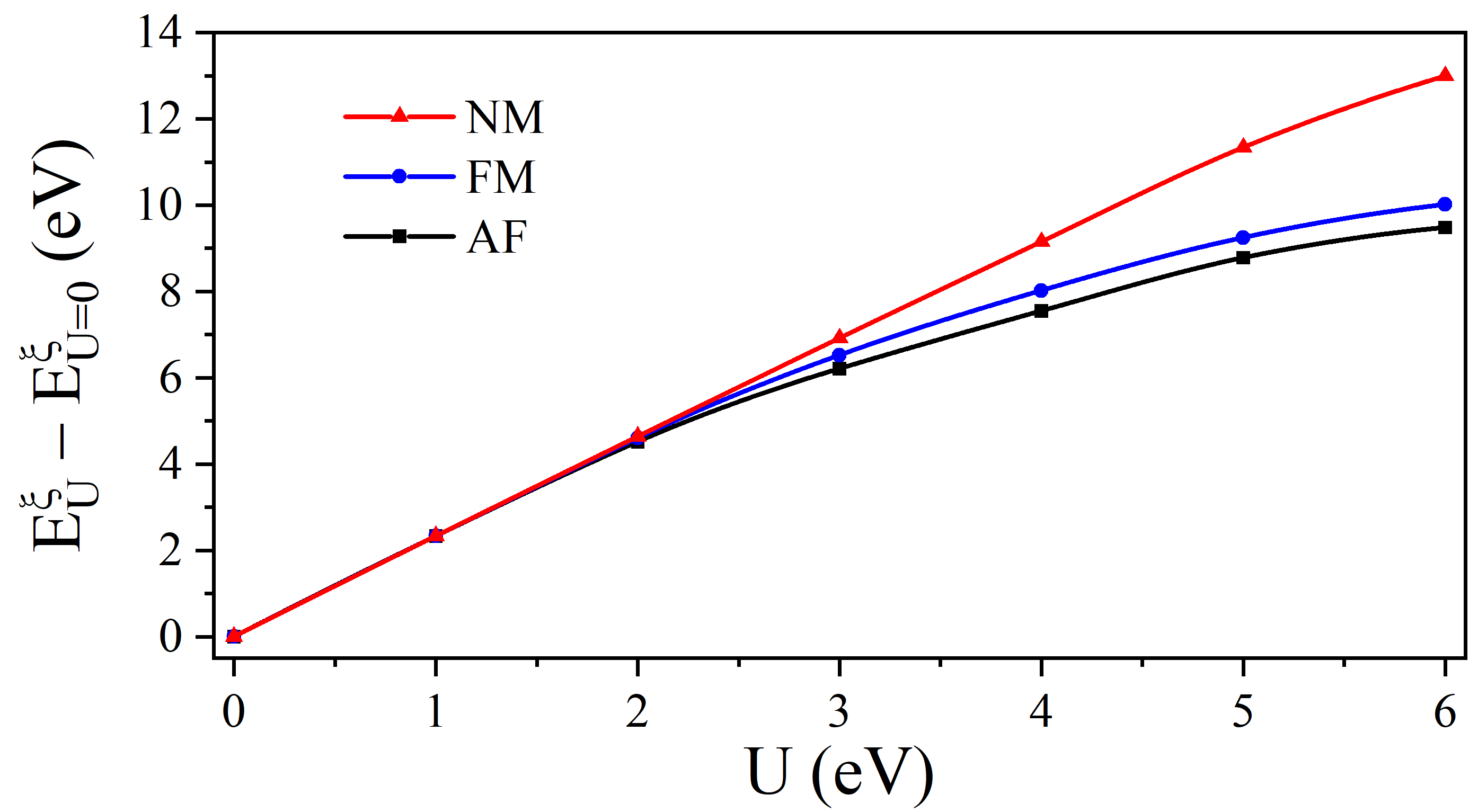} 
\caption{Total energy as a function of the U-Hubbard term. All energies were referenced to the 
energy of the same non-magnetic solution (U=0). $\xi$ labels the magnetic phase (AF, FM or NM)
Black dots represents the antiferromagnetic solution (AF), blue dots are for the ferromagnetic solution (FM), and red dots represents the non-magnetic solution (NM).}
\label{Fig:E-U}
\end{figure} 

For values of U-Hubbard different from zero, the magnetic solutions are energetically more stable than the non-magnetic solution. Despite the differences in the value of U used in VASP or QE in both codes, the AF state is the ground state. 
Figure \ref{Fig:E-U} shows the evolution of the stability order as a function of the 
U-Hubbard parameter. For small values of U-Hubbard, the magnetic moment of vanadium 
atoms tends to zero. For $U=0$ eV, the non-magnetic solutions are dominant; for 
$U = 1.0$ eV, the magnetic moment per vanadium atom is less than one Bohr magneton 
$1\, m_B$; and for $U > 4$ eV the magnetic moment is higher than $2\, m_B$. 
The slight difference between NM/FM/AF solutions can be understood by the low magnetic 
moment observed for solutions with $U<2$ eV.
Although V$_2$C monolayers have been synthesized recently using different 
experimental techniques\cite{wu2020synthesis,shekhirev2021characterization}, 
to the best of our knowledge, the antiferromagnetic character of the monolayer 
has not been tested or refuted yet.

As an additional comment to fig. \ref{Fig:E-U}, we would like to note that when long-range van der Waals interactions are not considered, although the stability order is the same, the non-magnetic (NM) and ferromagnetic (FM) solutions are closer in energy. For instance, for $U = 6$ eV, the difference in energy $\lvert E_{FM}-E_{NM} \rvert $ with van der Waals interaction is $\sim 3$ eV, and the same difference without van der Waals interaction is only $0.8$ eV. We hypothesize that such an effect may be due to the small changes in structure induced by the van der Waals forces, but further analysis is beyond the scope of this manuscript.

In Table \ref{tab:struc} we summarize the main structural parameters for the two 
cristallographic structures studied, the 1T ($1\times1$) and the $ 4 \times  4$ 
supercell for the different magnetic phases. 
The $ {4} \times  {4}$ lattice constant is divided by 4 for better comparison.
The AF solutions has the highest lattice constant (a$_0$); but the layer height ($\Delta_z$)  is the lowest. 
Finite-size effects can be seen in the structural parameters of the ferromagnetic (FM) solution; the variation in height could by enough to be experimentally measured.

\begin{table}[h!]
    \centering
\begin{tabular}{|c||c|c|c|c|c|}
\hline 
  & a$_0$  &  $\Delta_z$  & d$_{V-V}$  & d$_{V-C}$     \\ 
  & (\AA) &    (\AA) & (\AA) &  (\AA)   \\ 
\hline 
\hline 
\bf{NM}     & 2.881 & 2.191 & 2.751 &  1.992  \\ 
U=0  & (2.871) & (2.176) & (2.742) &  (1.988) \\ 
\hline 
\bf{AF}         & 3.548 & 1.817 & 2.738 &  2.241 \\ 
 U$\neq$0  & (3.548) & (1.816) & (2.738) &  (2.241) \\ 
\hline 
FM          & 3.265 & 2.188/2.203 & 2.895/2.921 &  2.195/2.218  \\ 
 U$\neq$0   &(3.252)& (2.181) & (2.903) & (2.204) \\ 
\hline 
\end{tabular} 
    \caption{Averaged structural parameters from the  $ {4} \times  {4}$ and --in parenthesis-- the   1T ($ {1} \times  {1}$) cell.
    a$_0$ is the scaled lattice constant --divided by 4 in the $ {4} \times  {4}$ supercell--, $\Delta_z$ is the layer height, d$_{V-V}$ is the V-V first inter-plane neighbors distance, and d$_{V-C}$ is the V-C first neighbors distance.}
    \label{tab:struc}
\end{table}

We found also, in agreement with previous 
reports\cite{Nyamdelger,ji2015different,hu2014investigations,champagne2018electronic}, 
that V$_2$C monolayer is metallic regardless of the value of U-Hubbard parameter employed.

\subsection{Lithium adsorption}
Next, we will explore the impact of the U-Hubbard parameter on the absorption energy 
of lithium at V$_2$C and whether there are changes in the variables associated with 
the description of the performance of a ion battery (open circuit voltage and storage capacity).
On the one hand, adsorption energies are measured experimentally using chemisorption 
experiments, and experimental surface characterization techniques are used to identify 
adsorption configurations\cite{bhola2017influence}. On the other hand, the 
characterization of V$_2$C-based lithium batteries can determine the open circuit voltage 
from which the binding energy for multiple lithium ions can be 
estimated\cite{yorulmaz2020systematical}.

\begin{figure}[!ht]
\includegraphics[clip,width=0.45\textwidth,angle=0,clip]{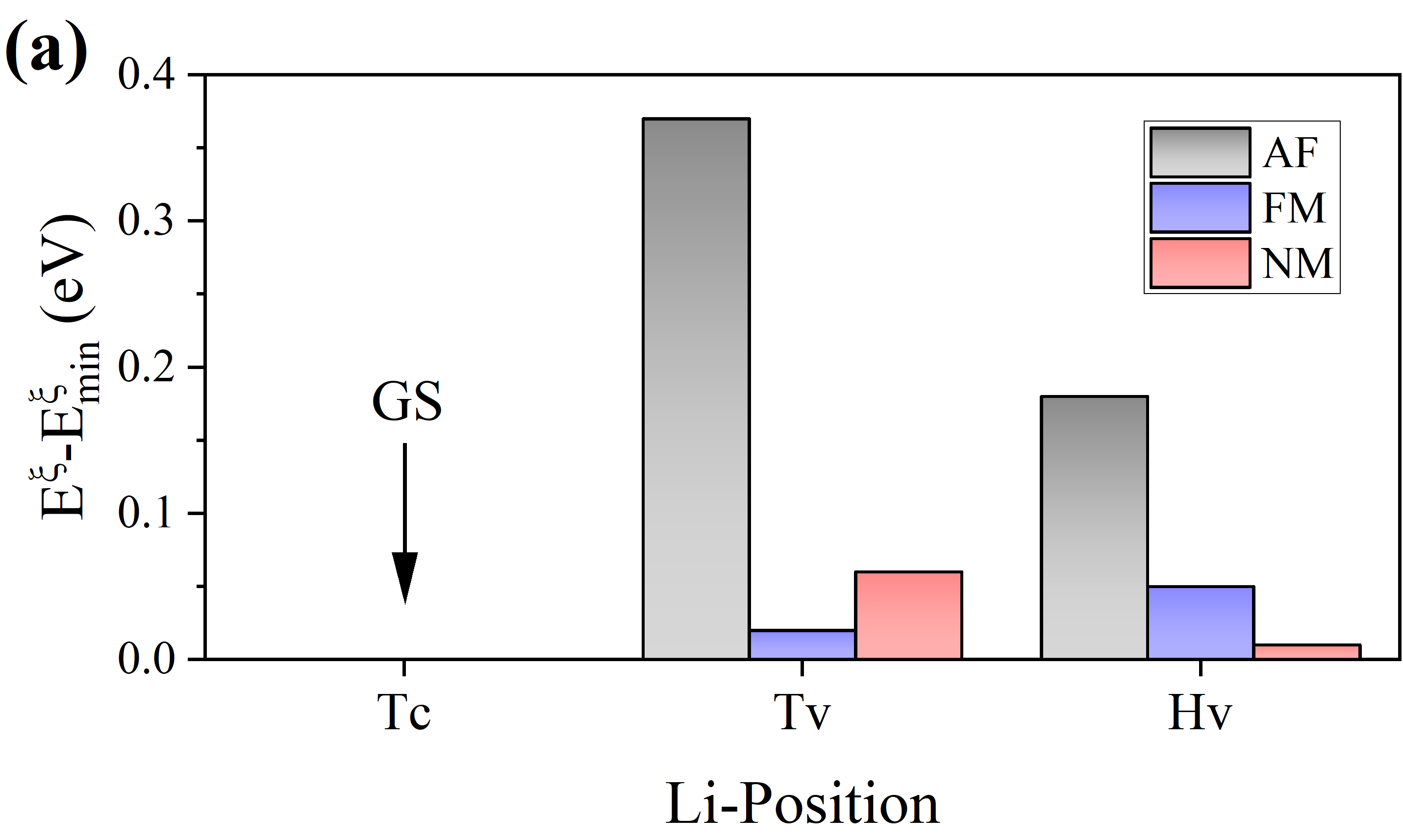} 
\includegraphics[clip,width=0.45\textwidth,angle=0,clip]{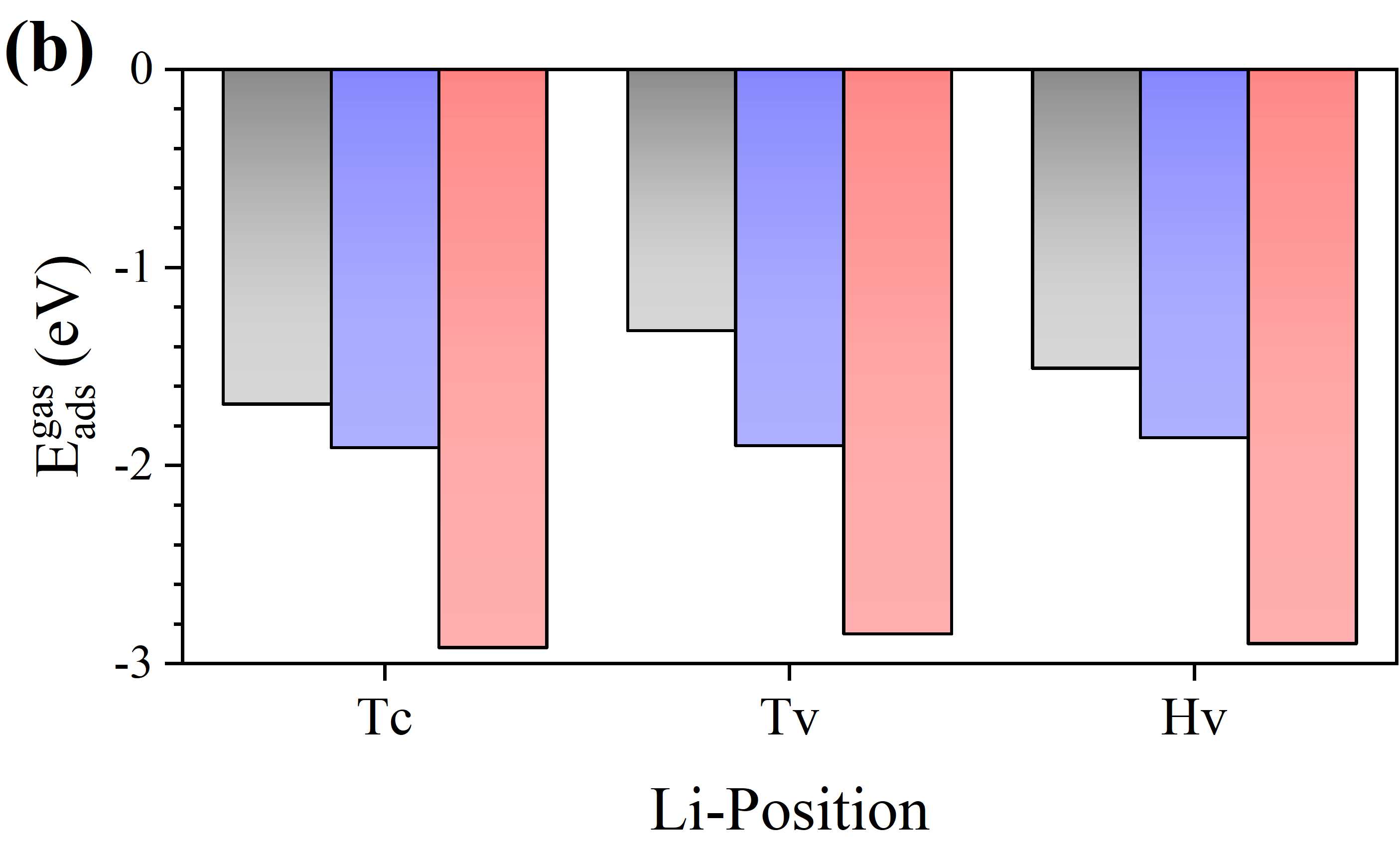} 
\caption{Lithium adsorption energies calculated with E$_{Li}$ in gas phase. 
In (a), the total energy relative to the ground state 
of each magnetic configuration, $E_{min}^{\xi}$, where $\xi$ labels the magnetic phase. 
In (b), the adsorption energy defined in eq. \ref{eq:Eads}.
Black for AF (U $\neq$ 0), blue for FM (U $\neq$ 0), and red for NM (U = 0).}
\label{Fig:1Li}
\end{figure}

We identify three preferred atomic positions (details in fig. \ref{Fig:scheme}): lithium on 
top of a carbon atom (T$_C$), lithium on top of a vanadium atom of the inner face (T$_V$), 
and lithium on top of a vanadium atom of the outer face (H$_V$).
We calculated for each adsorption site, the total energy of the configuration including the Li-ion and for every possible magnetic phase: 
NM (U = 0 eV), AF (U $\neq$ 0), and FM (U$\neq$ 0).
The preferred adsorption site, no matter the magnetic phase is the T$_C$ (Li on top of C).
In Fig. \ref{Fig:1Li}(a), we represent the energy difference between the different adsorption sites for each magnetic phase.  We define $E_{min}^{\xi}$ as the energy for each phase $\xi$ with a Li ion at the T$_C$ adsorption site and then separately compare the energy of the system for the other two adsorption sites for each phase ($E^{\xi}-E_{min}^{\xi}$).
The height of these bars indicates the barrier a Li ion should face when migrating from one adsorption site to the other. Remarkably, the difference in adsorption energies between the different meta-stable sites is the largest for the AF case, indicating higher mobility barriers.
For the other two phases, the difference in energy is more negligible, although the solutions have higher energy than AF.

In Fig. \ref{Fig:1Li}(b) we show the adsorption energy for each magnetic phase and for each site. This value indicates the feasibility of the V$_2$C monolayer as a battery, the smaller(more negative) the value the more difficult it is to extract the Li-ion from the system, the optimal value should be around 1 eV.
The adsorption energies is calculated by using eq. \ref{eq:Eads}. All calculations were performed in the 4x4 cell and Li in gas phase was used for the energy of the Li-ion, $E_{Li}$.

The adsorption energy also depends on the magnetic phase of the material. The AF configuration (U $\neq$ 0) has an adsorption energy of $E_{ads} =-1.5$ eV, followed by the FM solution (U $\neq$ 0) with $E_{ads} = -1.7$ eV and finally the NM solution (U = 0) with $E_{ads} = -2.7$ eV. If the system were non-magnetic this value of adsorption energy would rule out the use of this material as an electrode providing an argument in favor of using U.

We also find that for the V$_2$C monolayer, when adding two Li-ions, the system prefers that both Li are on opposite faces and bonded to the same carbon atom\cite{Nyandelger_2020}.
When we consider Lithium adsorption by both sides, it is energetically more favorable than the same side scenario\cite{Nyandelger_2020}. The adsorption energy is 56.3 meV higher in the NM case (U=0) and 89.7 meV higher in the AF case (U$\neq0$).

\subsection{Lithium-Battery capabilities}
Theoretical and experimental characterization of the V$_2$C monolayer as a 
potential electrode for lithium battery and the impact of simulation parameters 
is performed through open-circuit voltage (OCV) and charge capacity. 
The OCV is used to analyze the electronic energy 
changes in the electrode materials and estimate the battery's state of charge. 
The charge capacity relates the composition of the electrode to the amount of 
charge it can store; the storage capacity of Li-ion strongly depends on 
the surface functional groups changing with successive charge processes\cite{wu2020synthesis}.

To calculate the OCV eq. \ref{eq:OCV} is used for different concentrations of Lithium adding Li atoms  to the $4 \times 4$ cell of V$_2$C in different configurations. Each concentration of Li corresponds to a specific capacity. 
To obtain OCV values, we only consider the adsorption of Li in the T$_C$ positions. In the low concentration or highly dilute regimen, the absorption energy of the Li-ions depends on whether they are absorbed on the same face or different faces of the V$_2$C and coupled to the same carbon atom; the latter configuration is slightly more favorable, being $\sim 5$\% lower in energy\cite{Nyamdelger}

Additionally, we chose the position of the lithium ions to maximize the initial distance between 
them in few different configurations. Then for each concentration, we choose the one with the lowest energy. 
The difference in the adsorption energy of the system NM (U=0) and AF (U$\neq$0) 
decreases as the number of lithium adsorbed on the surface increases as can be seen from Fig. \ref{eq:OCV}(a).

It is worth noticing that when several Li-ions are added to the system the convergence of the calculations is accelerated when Li atoms bonded to the same carbon are not vertically aligned, i.e., the Li-C-Li angle deviates from 180. Therefore, we introduced slight random variations in 
the x and y components of the Li atom positions ($\Delta_{max} = \pm 0.1 $ \AA) away from carbon position and then start the relaxation procedure.

\begin{figure}[!ht]
\includegraphics[clip,width=0.45\textwidth,angle=0,clip]{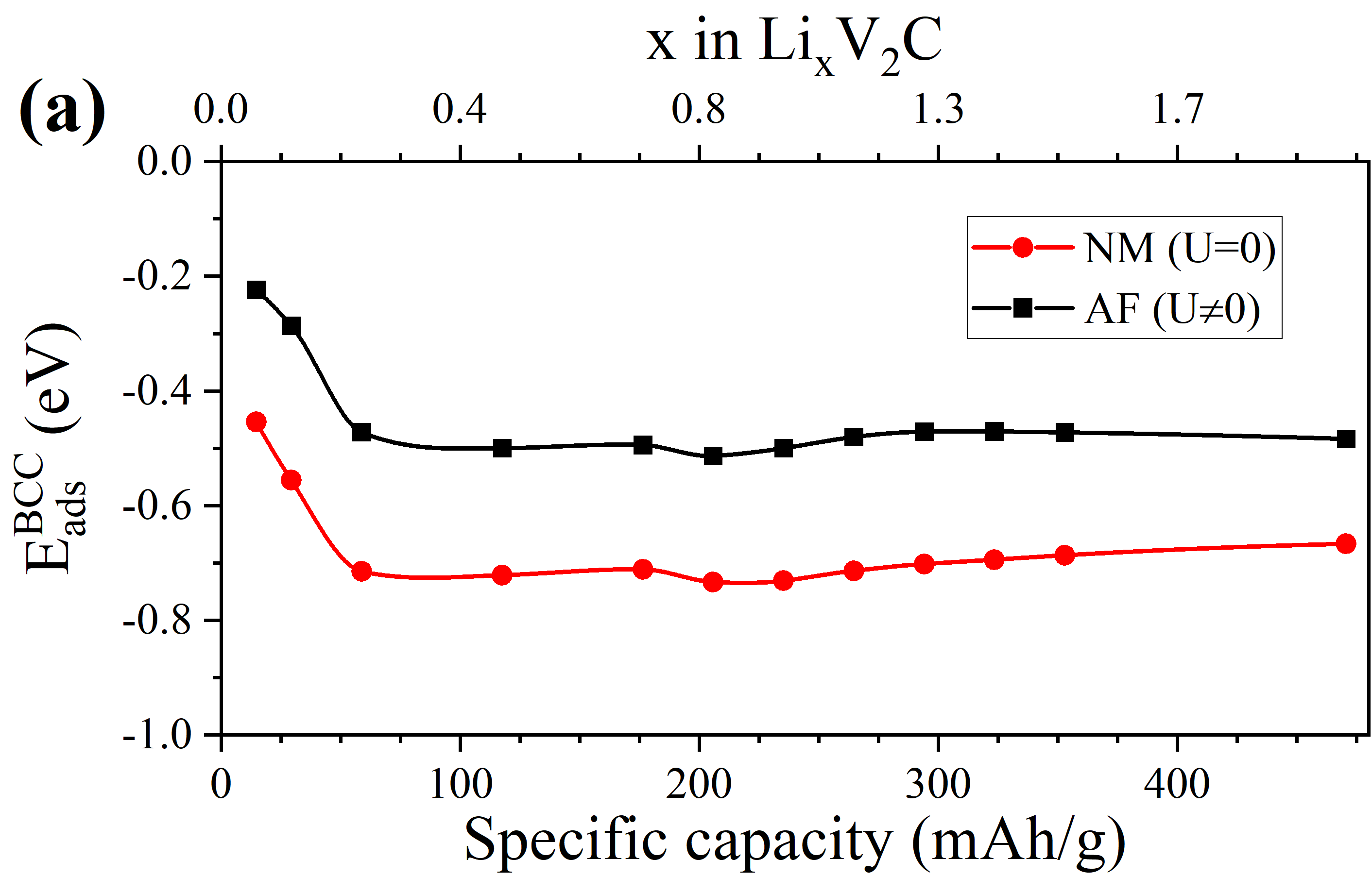}
\includegraphics[clip,width=0.45\textwidth,angle=0,clip]{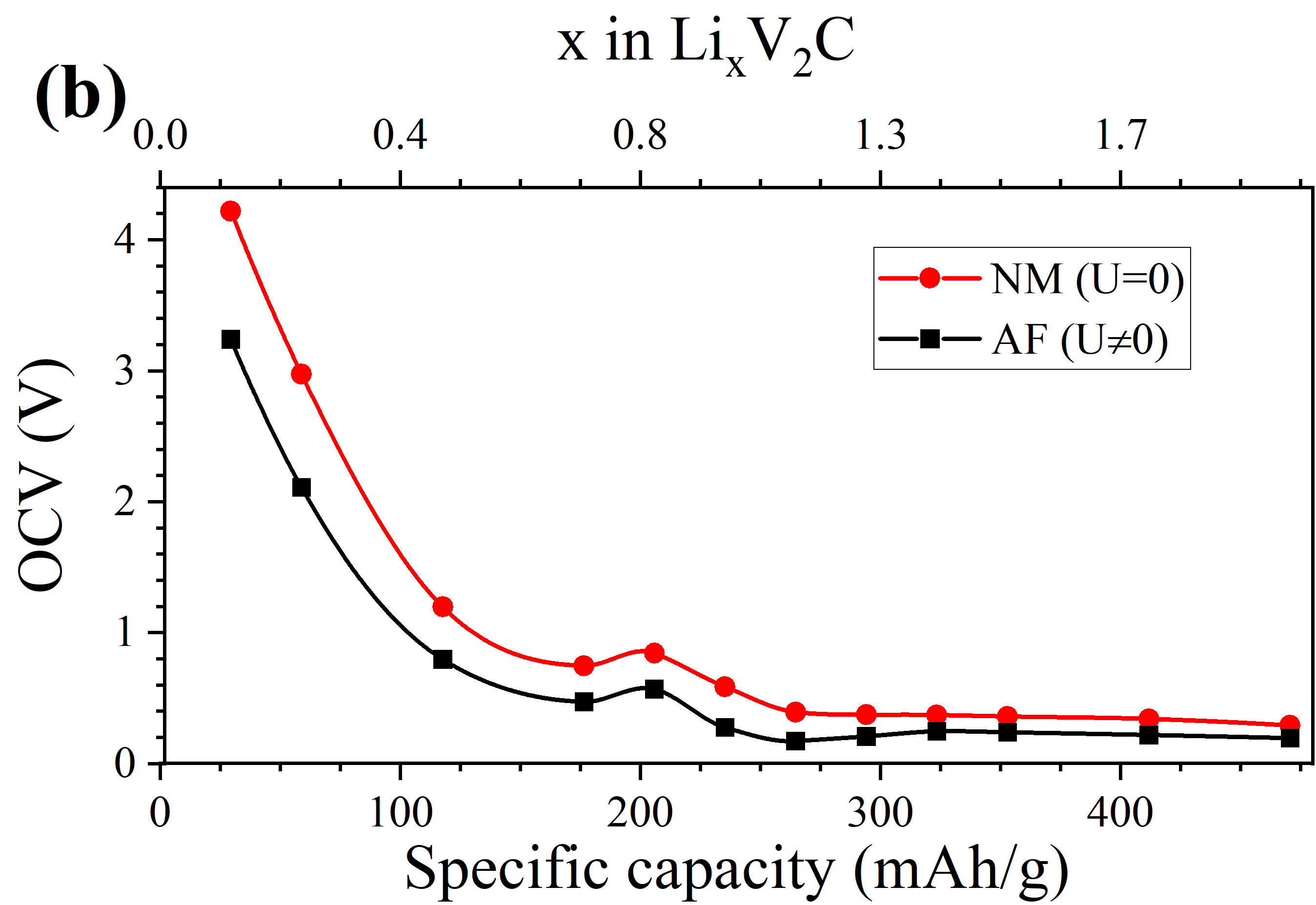} 
\caption{ Adsorption energy (a) and OCV (b) as a function of the Lithium concentration. Black lines for AF (U$ \neq $0) phase and red for NM (U = 0).
The adsorption energy and 
OCV values are calculated by using eq. \ref{eq:Eads}  and \ref{eq:OCV} respectively, in both cases we consider E$_{Li}$, the energy of a Li atom in a BCC crystal.}
\label{Fig:OCV}
\end{figure} 

The theoretical OCV values as a function of the concentration of lithium atoms in the V$_2$C 
monolayer are shown in fig. \ref{Fig:OCV}. 
The value of the adsorption and formation energy depends on considering the 
lithium in a crystalline or a gaseous state, the OCV 
also depends on the chemical origin of the Li-ion, as we mentioned in section \ref{metod},
we consider E$_{Li}$ the energy of a Li atom in a BCC crystal.

Note that the full stoichiometry corresponds to a specific capacity of up to 471 mAh/g. 
This theoretical specific capacity is in the range of the first cycle measurements of 
specific capacities of Li-ions batteries, with values around 500 mAh/g,  decreasing 
in the following charge cycles\cite{wu2020synthesis}.
Remarkably, our theoretical calculations reveal the same small bump in the OCV curve near 200 mAh/g as in the experimental measurements of ref.~\citen{wu2020synthesis}. This point correspond to a 14 lithium atoms concentration in the $ {4}\times {4}$ supercell, which in the 
adsorption-energy curve (fig. \ref{Fig:OCV}(a))can be identified as a change in the trend, reaching a  minimum in the curve. 
Such change in the curvature appears to be associated with a change in the crystallographic structure. Any couple of lithium atoms bonded to the same carbon atom below $x=0.7$ (Li$_x$V$_2$C) were vertically aligned; as the concentration increases, they prefer to move away from the vertical position. This shift causes some modifications in the position of Vanadium atoms, and as a result, the adsorption energy per Li atom is lowered, as we show in fig. \ref{Fig:OCV}(a). Further experimental measurements are needed to confirm this unusual behavior, although the same phenomena have been already observed in the two-sided Li adsorption in MoS$_2$ monolayers, which for a critical Li-concentration, stabilize a distorted structural 
phase\cite{nasr2015structural}.



\section{Final remarks}
In this work, we study the influence of the magnetic phase on the properties of the 
monolayer MXene V$_2$C as an electrode in lithium-ion-based batteries. We found that 
a single layer of V$_2$C behaves as an antiferromagnetic material and our results 
show that the Li-V$_2$C reaction profile strongly depends 
on the magnetic phase of the system and consequently on the election 
of the U-Hubbard parameter.
Such dependence on magnetic parameters and phases leads to changes in the quantities 
associated with battery performance, like adsorption energies and OCV curve. In that sense, 
from measurable properties, the magnetic state of the V$_2$C could be 
inferred and even used as a indirect validation of the U-Hubbard parameter 
determined by the linear response method.
The differences in the adsorption energy of a Li atom between the non-magnetic phase 
and the AF can be larger than $1$ eV; the OCV curves also differ in value even though 
they have the same trend.

Based on experimental results, we expect that in the evaluation of MXenes and specifically 
of V$_2$C as a battery electrode, it is necessary to consider the electronic correlation 
to obtain adequate results of the electrochemical parameters of the material. 
Surprisingly, the performance of a material in a battery can give us information about 
the magnetic phase and the range of the U-Hubbard parameter used to model the 
electronic correlation effects.

\section*{acknowledgment}
JWG acknowledges financial support from ANID-FONDECYT: Iniciaci\'on en Investigaci\'on 2019 grant N. 11190934 (Chile) and USM complementary research grant year 2021.
ESM acknowledges financial support from ANID-FONDECYT 1170921.

\bibliography{v2c_bib}
\end{document}